\documentclass[letterpaper, 10 pt, conference]{ieeeconf}  

\IEEEoverridecommandlockouts                              

\usepackage{color}
\usepackage{amssymb}
\usepackage{graphicx}
\usepackage{amsmath}
\usepackage{cite}
\usepackage{multirow}
\usepackage{cuted}
\usepackage{soul} 
\usepackage[table]{xcolor}
\usepackage{hyperref}

\title{\LARGE \bf
SurGSplat: Progressive Geometry-Constrained Gaussian Splatting for Surgical Scene Reconstruction
}

\begin{document}
\begin{sloppypar}

\author{Yuchao Zheng$^{1}$, Jianing Zhang$^{2}$, Guochen Ning$^{3}$ and Hongen Liao$^{1}$ \IEEEmembership{Senior Member, IEEE} 
\thanks{This work was supported by the National Natural Science Foundation of China (U22A2051, 82027807), National Key Research and Development Program of China (2022YFC2405200), Beijing Municipal Natural Science Foundation (7212202), Institute for Intelligent Healthcare, Tsinghua University (2022ZLB001), Tsinghua-Foshan Innovation Special Fund (2021THFS0104), and Young Elite Scientists Sponsorship Program by CAST (2023QNRC001). (Corresponding author: Guochen Ning)}
\thanks{
$^{1}$Y. Zheng is with the School of Biomedical Engineering, Tsinghua University, Beijing 100084, China 
     {\tt\small zhengyc23@mails.tsinghua.edu.cn}.%
$^{2}$J. Zhang is with the School of Information Science and Technology, Fudan University, Shanghai 200433, China
        {\tt\small 22110720080@m.fudan.edu.cn}.%
$^{3}$G. Ning is with the School of Clinical Medicine, Tsinghua University, Beijing 100084, China
        {\tt\small ningguochen@tsinghua.edu.cn}.%
$^{1}$H. Liao is with the School of Biomedical Engineering, Tsinghua University, Beijing 100084, China
        {\tt\small liao@tsinghua.edu.cn}.}%
}

\maketitle
\thispagestyle{empty}
\pagestyle{empty}

\begin{strip}
\begin{minipage}{\textwidth}
    \vspace{-95pt}
    \centering
    \includegraphics[width=0.95\textwidth]{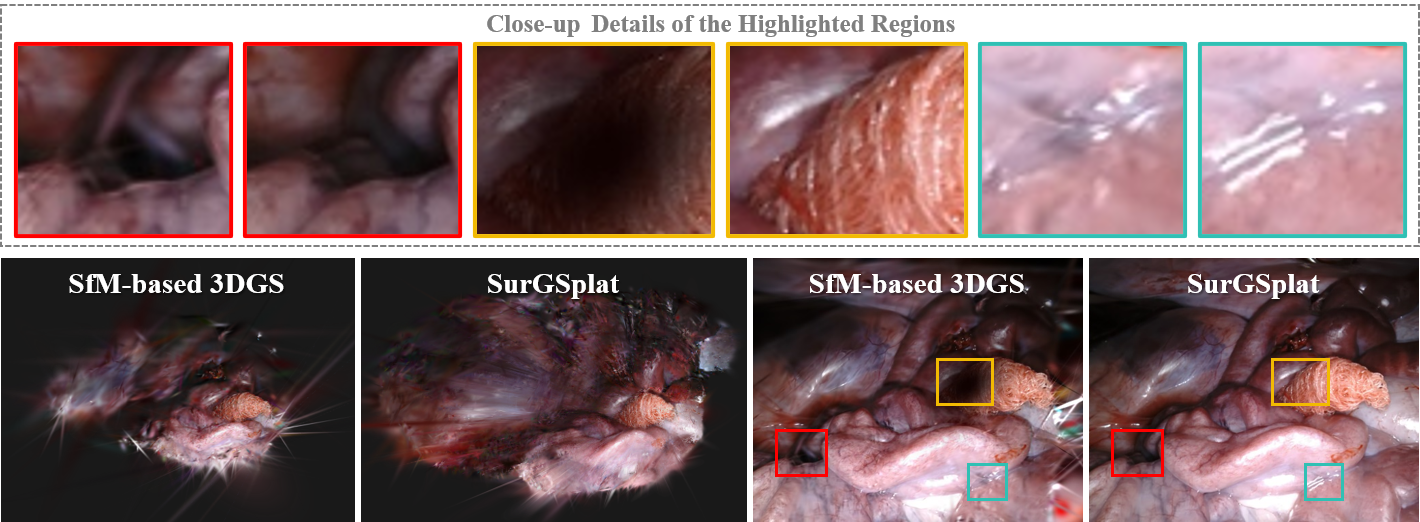}
    {\small \textbf{Teaser:} Comparison of 3D reconstruction and novel view synthesis between SfM-based 3DGS and our \textbf{SurGSplat} method, demonstrating SurGSplat’s ability to achieve more realistic and high-fidelity reconstructions \textbf{without relying on SfM poses}, effectively addressing challenges in surgical scenarios.}
    \label{teaser}
\end{minipage}
\end{strip}

\begin{abstract}

Intraoperative navigation relies heavily on precise 3D reconstruction to ensure accuracy and safety during surgical procedures. However, endoscopic scenarios present unique challenges, including sparse features and inconsistent lighting, which render many existing Structure-from-Motion (SfM)-based methods inadequate and prone to reconstruction failure. To mitigate these constraints, we propose SurGSplat, a novel paradigm designed to progressively refine 3D Gaussian Splatting (3DGS) through the integration of geometric constraints. By enabling the detailed reconstruction of vascular structures and other critical features, SurGSplat provides surgeons with enhanced visual clarity, facilitating precise intraoperative decision-making. Experimental evaluations demonstrate that SurGSplat achieves superior performance in both novel view synthesis (NVS) and pose estimation accuracy, establishing it as a high-fidelity and efficient solution for surgical scene reconstruction. More information and results can be found on the page https://surgsplat.github.io/.

\end{abstract}

\section{INTRODUCTION}

Accurate intraoperative scene reconstruction serves as a cornerstone for precise control of surgical robots and effective navigation in minimally invasive procedures \cite{freedygs2024}. Traditional endoscopic 3D reconstruction methods rely on explicit geometric representations (e.g., point cloud, mesh) \cite{mesh2024}, which often yield low-resolution models with insufficient details. Moreover, these methods often rely on preoperative radiological scans or external depth sensors, which not only increase the complexity of preoperative preparations but also reduce their flexibility and accuracy during real-time applications \cite{ct2020}.

\begin{figure*}[thpb]
    \vspace{3pt}
    \centering
    \includegraphics[width=0.95\linewidth, height=0.9\textheight, keepaspectratio]{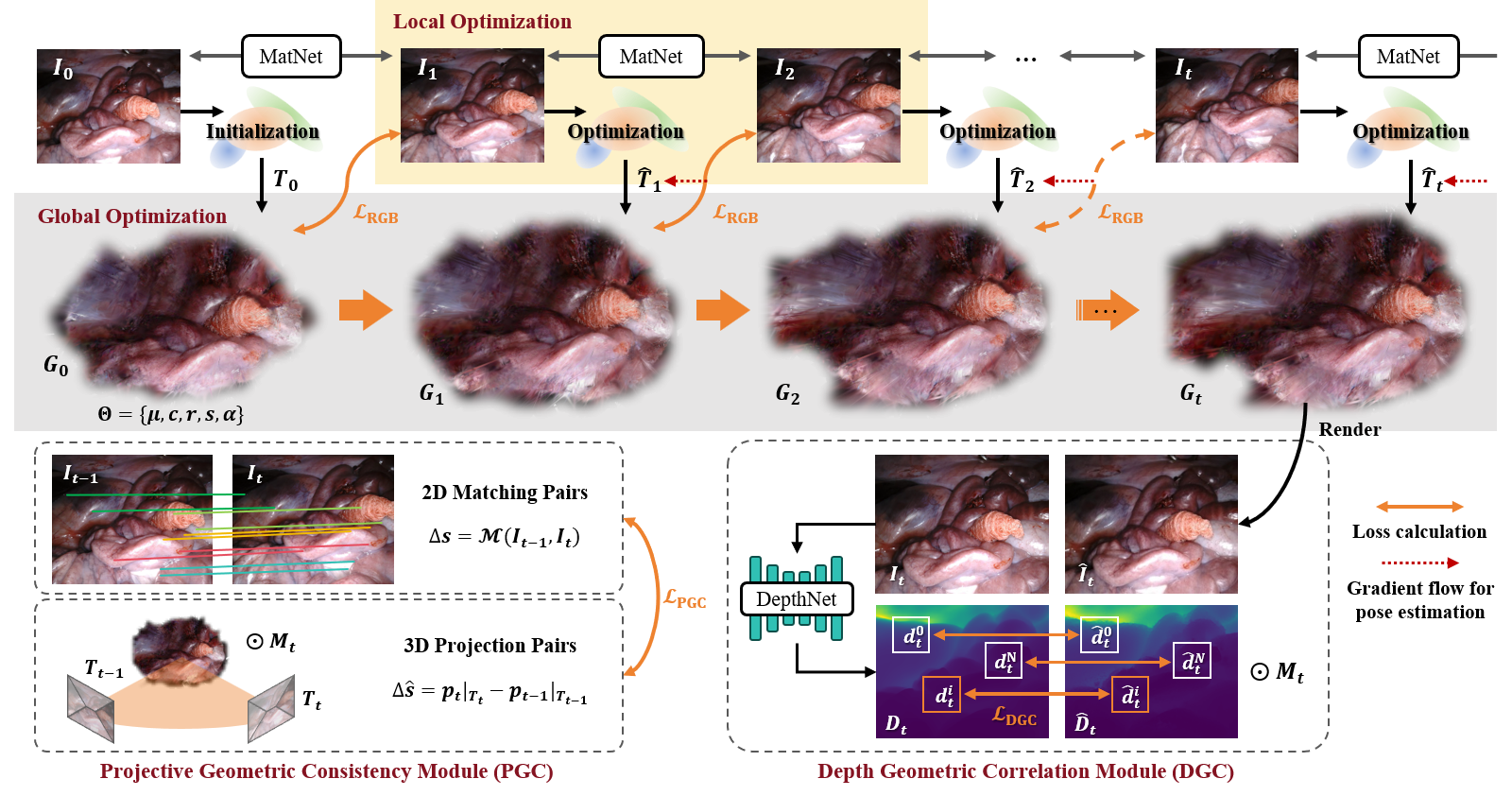}
    \caption{Pipeline of \textbf{SurGSplat} for high-fidelity surgical scene reconstruction. Starting with an initial depth map generated from a single image $I_0$, our method iteratively optimizes camera poses and global 3D Gaussian representations. Each frame is processed by constructing a local 3D Gaussian set and optimizing the relative pose via neural rendering. The accumulated poses are used to refine the global 3D Gaussians based on geometric consistency losses.}
    \label{Fig:1}
\end{figure*}


The introduction of Neural Radiance Fields (NeRF) \cite{nerf2020} has revolutionized medical image reconstruction by generating realistic novel-view images through implicit representations, excelling in scenes with complex lighting conditions. However, this implicit representation leads to inherent challenges, such as slower training speeds and difficulty in capturing high-frequency details, such as capillaries. This presents substantial obstacles for downstream tasks like lesion detection, augmented reality (AR) \cite{ar2018}, and intraoperative navigation. In recent years, the 3D Gaussian Splatting (3DGS) \cite{3dgs2023} method has made significant strides in both 3D reconstruction and rendering efficiency \cite{2dgs2024, pgsr2024, gbr2024}. By combining the advantages of implicit representations and explicit point cloud rendering, 3DGS models scene points using Gaussian distributions. This approach enables high-precision 3D modeling with less time consumption and reduced reliance on pretrained data, offering superior real-time performance. Despite these advantages, both NeRF and 3DGS methods generally depend on accurate camera poses, with 3DGS requiring an initial point cloud typically generated through Structure-from-Motion (SfM) methods (e.g., COLMAP) \cite{colmap2016mvs, colmap2016sfm}. However, in endoscopic environments, the sparse textures, intricate anatomical structures, and photometric inconsistencies (e.g., reflections, non-Lambertian surfaces, and variable lighting) limit the number of reliable feature points for SfM-based initialization, frequently resulting in overly sparse point clouds or reconstruction failures \cite{challenge2024, challenges22024}.

To reduce the dependency on SfM initialization, researchers have proposed several innovative NeRF-based methods that cleverly integrate pose estimation into the reconstruction process \cite{inerf2021, nerfmm2021, barf2021, nopenerf2023}. For instance, i-NeRF \cite{inerf2021} predicts poses using a pretrained model, NeRFmm \cite{nerfmm2021} jointly optimizes both camera pose and NeRF parameters, and NoPe-NeRF \cite{nopenerf2023} incorporates geometric priors to optimize camera poses. However, these methods typically rely on strict scene constraints or precise initial poses, and they exhibit limitations in optimization efficiency. To address these issues, SfM-free methods based on 3DGS have been introduced, such as CF-3DGS \cite{cf3dgs2024}. This approach leverages the temporal continuity of the video, constructing local Gaussian point clouds frame by frame and globally fusing them, effectively eliminating the need for SfM preprocessing and significantly improving the real-time performance and robustness of 3D reconstruction. However, this method lacks geometric constraints, which may result in a decrease in the detail and accuracy of the reconstruction in complex scenes, such as endoscopic environments.

In this paper, we propose \textbf{SurGSplat} to realize high-fidelity \textbf{Sur}gical scene reconstruction based on 3D \textbf{G}aussian \textbf{Splat}ting while ensuring high-quality rendering. It incorporates geometric constraints into the progressive optimization of global 3D Gaussians using the temporal continuity of endoscopic video frames, addressing challenges such as sparse textures and photometric inconsistencies. The overall pipeline is illustrated in Fig.\ref{Fig:1}. Firstly, it generates an initial depth map from a single view to initialize the 3D Gaussians, followed by the iterative optimization of relative camera poses and the global Gaussian representation. For each frame $t$, a local 3D Gaussian set is constructed based on the previous frame $t-1$, and an affine transformation $A_t$ is optimized through neural rendering to estimate the relative pose $T_{t-1\to t}$. Subsequently, the information from the current frame is integrated into the global 3D Gaussians via the accumulated relative pose $T_{1\to t}$. Specifically, during the 3D Gaussians optimization, visual loss, Projection Geometric Consistency Loss (PGC), and Depth Geometric Correlation Loss (DGC) are employed as supervision, ensuring accurate alignment and refinement of the 3D structure. These constraints facilitate the iterative accumulation and optimization of a robust 3D scene reconstruction tailored for complex endoscopic environments. Our contributions can be summarized as follows:

\begin{enumerate}
    \item \textbf{SfM-Free 3D Reconstruction.} Developed an SfM-free method for 3DGS-based reconstruction in endoscopic scenes, removing the reliance on camera pose estimation traditionally required by SfM methods.
    \item \textbf{Geometric Constraints.} Incorporated depth and projective geometric-constrained losses to improve alignment accuracy and robustness in 3D reconstruction under complex structures.
    \item \textbf{Progressive Optimization.} Utilized temporal continuity in endoscopic videos to progressively optimize 3D Gaussians, enabling efficient and accurate scene reconstruction.
\end{enumerate}

\section{METHOD}
Given a sequence of unposed endoscopic images $\{I_t\}, t\in(0, N-1)$ with camera intrinsic $K$, our objective is to restore the camera trajectory $\{T_t\}$ and incrementally reconstruct the surgical scene $\{G_t\}$. To this end, we propose \textbf{SurGSplat}, a novel paradigm that simultaneously optimizes 3D Gaussians and camera poses while integrating geometric constraints tailored for surgical environments. Our method begins with a review of the 3DGS framework (Sec.\ref{Sec.2.1}), followed by a local-to-global approach for scene reconstruction (Sec.\ref{Sec.2.2}). This includes initializing 3D Gaussians from a single view, estimating relative camera poses, and progressively expanding global 3D Gaussians. Finally, we introduce geometry-constrained optimization (Sec.\ref{Sec.2.3}), leveraging projective and depth geometric losses to refine the reconstruction and ensure structural coherence.

\subsection{Preliminary}\label{Sec.2.1}

3D Gaussian Splatting (3DGS) models a scene as a collection of explicit 3D Gaussian ellipsoids initialized from sparse point clouds and calibrated camera poses obtained through SfM. Each Gaussian is parameterized by $\Theta^i = \{\mu^i, c^i, r^i, s^i, \alpha^i\}$, where \(\mu\in\mathbb{R}^3\) denotes the center, \(c\in\mathbb{R}^3\) represents color coefficients, \(r\in\mathbb{R}^4\) is the quaternion rotation, \(s\in\mathbb{R}^3\) specifies the scale, and \(\alpha\in\mathbb{R}\) is the opacity. The density function of each Gaussian is given by : 
\begin{equation}
    G^i(x) = e^{-\frac{1}{2}(x-\mu^i)^\top {\Sigma^i}^{-1}(x-\mu^i)}.
\end{equation}

The covariance matrix of a 3D Gaussian, $\Sigma^i = R^i S^i {S^i}^\top {R^i}^\top,$ encodes the ellipsoidal shape of the Gaussian. Here, \(R\) represents the orthonormal basis of the ellipsoid's principal axes, while \(S\) defines the scaling factors along each axis, determining the size of the ellipsoid. To facilitate optimization, the 3D Gaussian scene is required to be projected onto the 2D plane corresponding to a specified camera pose \( T_t \). In the camera coordinate system, the covariance matrix \( \Sigma \) is expressed as $\Sigma_t^{\text{2D}} = J T_t \Sigma T_t^\top J^\top,$ where \( J \) represents the Jacobian matrix derived from the affine approximation of the projective transformation. The final color \(\hat{C}\) and depth \(\hat{D}\) for each pixel is computed via alpha blending of the projected Gaussians:
\begin{equation}
    \hat{C} = \sum_{i=0}^N c^i \alpha^i \prod_{j<i}(1-\alpha^j), \quad \hat{D} = \sum_{i=0}^N d^i \alpha^i \prod_{j<i}(1-\alpha^j),
\end{equation}
where \(c^i, \alpha^i\) are the color and opacity of the \(i\)-th Gaussian, and \(d^i\) is its depth, determined from the \(z\)-coordinate of \(\mu\) in the camera space. This explicit and differentiable representation allows for efficient optimization of both geometry and photometry.

\subsection{Local Pose Estimation and Global 3DGS Growing}\label{Sec.2.2}

\subsubsection{3DGS Initialization from a Single View}

To tackle the challenges of intraoperative scene reconstruction, we initialize the process using the first frame \(I_0\). Based on that, we use a \textit{DepthNet} to estimate the monocular depth map \(D_0\), offering strong geometric cues. Assuming an identity camera pose \(T_0 = \mathbf{I}\) and leveraging the camera intrinsics \(K\), the depth map \(D_0\) is projected into 3D space to establish the initial 3D Gaussian set \(G_0\). Subsequently, the attributes \(\Theta_0 = \{\mu_0, c_0, r_0, s_0, \alpha_0\}\) are refined by minimizing the photometric loss between the rendered image \(\mathcal{R}(G_0)\) and the frame \(I_0\). The optimization is defined as:  
\begin{equation}
    G_0^* = \underset{c_0, r_0, s_0, \alpha_0}{\arg \min} \mathcal{L}_{\text{RGB}}(\mathcal{R}(G_0, T_0), I_0),
\end{equation}
where the photometric loss is a weighted combination of the L1 and D-SSIM terms:  
\begin{equation}
    \mathcal{L}_{\text{RGB}} = (1 - \lambda_{rgb}) \mathcal{L}_1 + \lambda_{rgb} \mathcal{L}_{\text{D-SSIM}}.
\end{equation}

This framework effectively harnesses the geometric information from the initial intraoperative frame, providing a robust basis for high-fidelity scene reconstruction in surgical environments.

\subsubsection{Relative Pose Estimation} 
To facilitate intraoperative scene reconstruction, we establish the relationship between the camera pose \(T_t\) and the rigid transformation of 3D Gaussian points \(A_t\). Specifically, given a set of 3D Gaussians with centers \(\mu\), their 2D projections \(\mu^{\text{2D}}\) under the camera pose \(T_t\) are expressed as:  
\begin{equation}
    \mu^{\text{2D}} = \frac{K T_t \mu}{\mathbf{v}_z^\top (T_t \mu)},
\end{equation}
where \(\mathbf{v}_z^\top (T_t \mu)\) extracts the depth component along the optical axis. Alternatively, the 2D projections can also be computed for a set of points \(\mu'\), rigidly transformed under an affine transformation \(A_t\):  
\begin{equation}
    \mu^{\text{2D}} = \frac{K \mathbb{P} \mu'}{\mathbf{v}_z^\top (\mathbb{P} \mu')}, \quad \mu' = A_t \mu,
\end{equation}
where \(\mathbb{P}\) represents the orthogonal projection direction. This equivalence reveals that estimating the camera pose \(T_t\) is mathematically analogous to determining the affine transformation \(A_t\) of 3D Gaussian kernels. Accordingly, the pose estimation task is cast as minimizing the photometric loss \(\mathcal{L}_{\text{RGB}}\) between the rendered view \(\mathcal{R}(A_t \odot G_t)\) and the subsequent frame \(I_{t+1}\):  
\begin{equation}
    A_t^* = \underset{A_t}{\arg\min} \ \mathcal{L}_{\text{RGB}}(\mathcal{R}(A_t \odot G_t), I_{t+1}).
\end{equation}

During this optimization, the attributes of the 3D Gaussian set \(\Theta_t\) remain fixed, effectively isolating the estimation of \(T_t\) from Gaussian deformation and pruning.

\subsubsection{Global Progressive Growth of 3DGS}

The local approach effectively estimates relative poses \(T_t\) and constructs Gaussian scenes for two-frame video segments. However, the narrow FOV, complex illumination, and non-rigid inter-frame motion make local scenes fail to fully cover the surgical environment. Besides, noise in relative pose estimation limits reconstruction precision and robustness. To address these challenges, we introduce a globally progressive optimization strategy that incrementally updates the global Gaussian scene, enabling efficient and rapid scene completion with each new frame.

This process is divided into four steps: \textbf{(a) Initialization.} For the \((t-1)\)-th frame \(I_{t-1}\), we initialize a set of 3D Gaussian points \(G_{t-1}\) with an orthogonal camera pose \(T_{t-1} = \mathbf{I}\) (identity transformation). This step provides the initial sparse scene representation. \textbf{(b) Sequential Updates.} Using the local Gaussians with the relative pose \(T_{t-1 \to t}\), the geometric relationship between frames is updated, progressively refining the global Gaussian scene \(G\). \textbf{(c) Global Optimization and Densification.} Gradient information derived from inter-frame parallax is leveraged to identify under-reconstructed regions, where Gaussian points are progressively added to achieve densification. Over time, as more frames are integrated, \(G\) evolves from a sparse representation to a comprehensive scene that covers the entire intraoperative environment. To further ensure accuracy, the geometric consistency loss is employed during optimization, which enforces frame-to-frame consistency and effectively mitigates errors caused by noise in \(T_t\) and non-rigid motion. Details can be found in Sec.\ref{Sec.2.3}. \textbf{(d) Continuous Growth.} Instead of terminating densification midway, our framework maintains continuous growth of \(G\) throughout the input sequence. This ensures that new frames dynamically expand the coverage and detail of the global scene, resulting in a complete and robust surgical scene reconstruction.

\subsection{Geometry-constrained 3DGS Optimization}\label{Sec.2.3}

\begin{figure}[ht]
    \centering
    \includegraphics[width=1.0\linewidth]{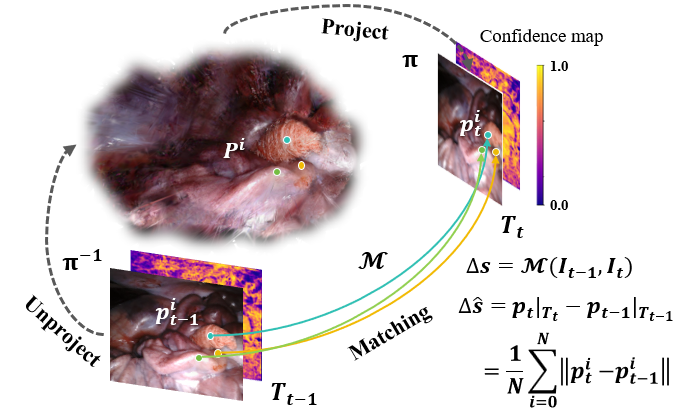}
    \caption{Principle of \textbf{Projection Geometric Consistency Loss (PGC)}. The matching distance $\Delta s = \mathcal{M}(I_{t-1}, I_t)$ serves as ground truth to optimize the projective distance $\Delta \hat{s}$ between corresponding 3D points projected from consecutive frames, ensuring geometric consistency across matched features.
}
    \label{Fig:2}
\end{figure}

The unique characteristics of endoscopic imagery, such as sparse textures and intricate structures, pose significant challenges for reliable feature matching, leading to frequent failures in conventional SfM approaches. To overcome these limitations and improve the fidelity of Gaussians, we propose two complementary geometric constraints: the Projection Geometric Consistency Loss (PGC) and the Depth Geometric Correlation Loss (DGC). Additionally, a depth correction mask $M_t$ is applied to clamp the rendered depth values that exceed a predefined threshold $k$, mitigating the impact of outliers on the optimization process. Combined with the proposed geometric constraints, this approach introduces robust supervisory signals that integrate 2D-3D geometric priors into Gaussian representations, effectively addressing the challenges in endoscopic environments.

\textbf{Projection Geometric Consistency Loss (PGC)} introduces a rigorous geometric constraint to ensure coherence between matched points in consecutive frames and their corresponding projections in 3D space. This formulation explicitly aligns 2D image features with their re-projected counterparts, thereby enhancing the fidelity and stability of the reconstructed Gaussian scene. The 2D matching distance can be defined as $\Delta s = \mathcal{M}(I_{t-1}, I_t)$, where $MatNet$ ($\mathcal{M}$) is a dedicated model designed for dense feature points matching between image pairs. As shown in Fig.\ref{Fig:2}, we introduce a projection flow to calculate the movement of matched points between consecutive frames by projecting 3D Gaussians from camera pose \(T_{t-1}\) to \(T_t\). Specifically, for a given pixel \(p^i_{t-1}\) in frame \(I_{t-1}\), its corresponding 3D position \(P^i_{t-1}\) in the Gaussian scene is obtained via back-projection using the rendered depth \(D_{t-1}\):
\begin{equation}
    P^i_{t-1} = \pi^{-1}(T_{t-1}, D_{t-1}(p^i_{t-1}), K),
\end{equation}
where \(\pi^{-1}\) denotes the back-projection operator, and \(D_{t-1}(p^i_{t-1})\) represents the depth value at pixel \(p^i_{t-1}\). The corresponding point \(p^i_t\) in frame \(I_t\) is computed by projecting \(P^i_{t-1}\) to the current frame's camera pose \(T_t\):
\begin{equation}
    p^i_t = \pi(T_t, P^i_{t-1}, K),
\end{equation}
where \(\pi\) is the projection operator from 3D space to the 2D image plane. Notably, to improve the reliability of point correspondences, points $p^i$ with confidence scores below 20\% were excluded from the projection loss calculation.

Assuming that the pose estimation \(T_{t}\) obtained from the previous local optimization step is accurate, for all \(N\) matched feature points \(p^i_{t-1}\) in frame \(I_{t-1}\) and their corresponding projections \(p^i_t\) in frame \(I_t\), the geometric discrepancy \(\Delta \hat{s}\) is defined as the mean Euclidean distance $\Delta \hat{s} = \frac{1}{N} \sum_{i=1}^N \left\| p^i_t - p^i_{t-1} \right\|$. The PGC loss is formulated as:

\begin{equation}
    \mathcal{L}_{\text{PGC}} = \left\|\Delta s - \Delta \hat{s} \right\|.
\end{equation}

\begin{figure*}[ht]
    \centering
    \includegraphics[width=0.98\linewidth, height=0.9\textheight, keepaspectratio]{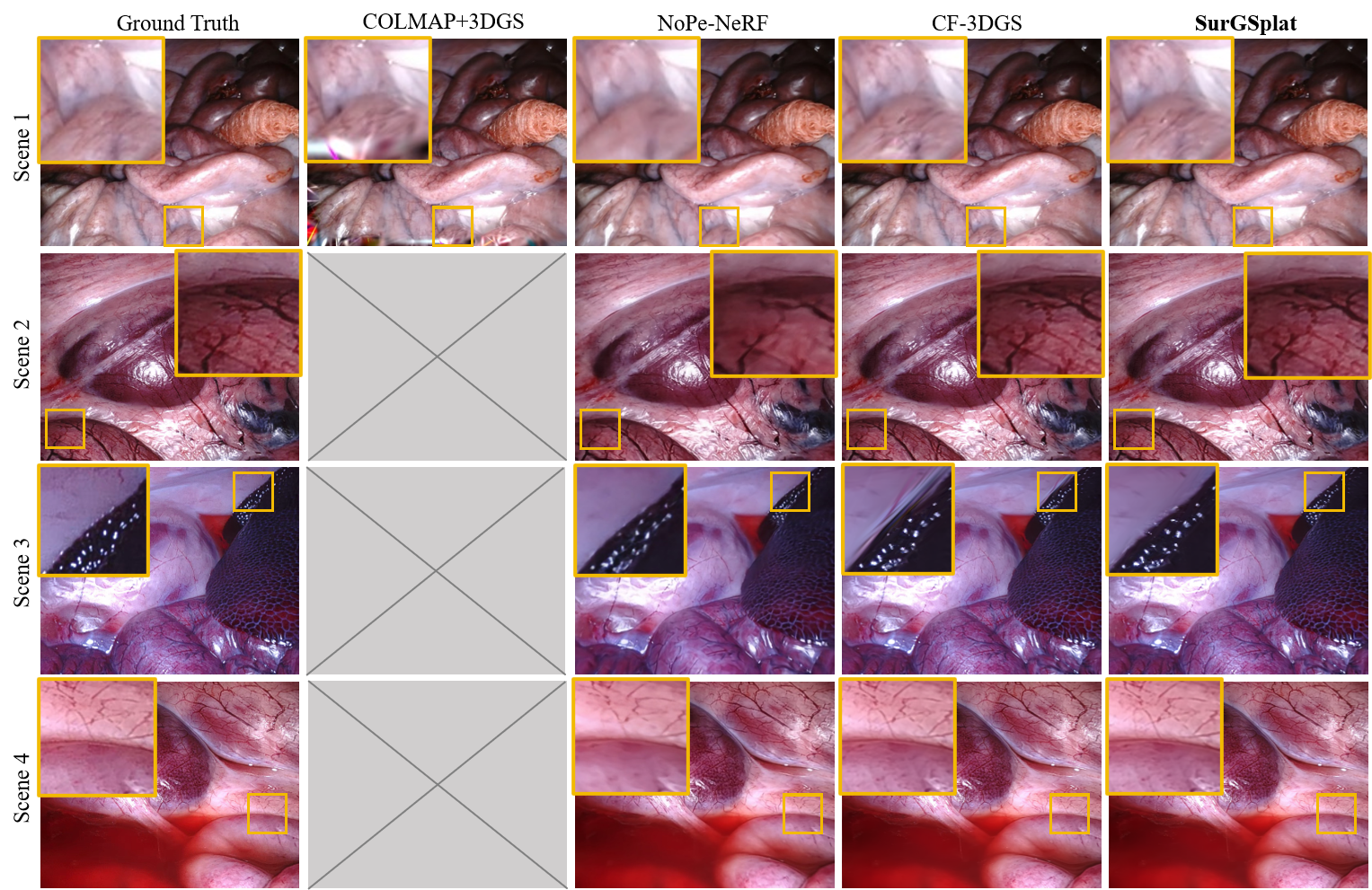}
    \caption{Rendering results for four representative scenes in the SCARED dataset. Notably, COLMAP only successfully reconstructed Scene 1, while failing on the other scenes.}
    \label{Fig:3}
\end{figure*}

\textbf{Depth Geometric Correlation Loss (DGC)} further incorporates depth information as a complementary geometric constraint. For each frame \(I_t\), a depth map \(D_t\) is generated using a pre-trained \textit{DepthNet}, while a corresponding depth map \(\hat{D}_t\) is rendered from the Gaussian scene based on the camera pose \(T_t\). To evaluate the geometric correlation between two depth maps, we compute a local Pearson correlation-based loss. By normalizing depth values within each patch, it ensures robustness to global scale variations, making the loss unaffected by differences in absolute depth. Furthermore, it penalizes deviations in relative depth trends, promoting fine-grained alignment of reconstructed geometries, which is essential for applications like surgical navigation. Specifically, \(N\) random patches $d_t^i$ and $\hat{d}_t^i$ of size \(a \times a\) are sampled from the ground truth $D_t$ and the rendered depth map $\hat{D}_t^i$, and the correlation within each patch is calculated and averaged as:
\begin{equation}
    \mathcal{L}_{\text{DGC}} = 1 - \frac{1}{N} \sum_{i=1}^N \frac{\text{Cov}(d_t^i, \hat{d}_t^i)}{\left( \sigma(d_t^i) + \epsilon \right) \cdot \left( \sigma(\hat{d}_t^i) + \epsilon \right)}.
\end{equation}
where $\text{Cov}$ represents the covariance, and $\sigma$ is the standard deviation. $\epsilon$ is introduced as a small positive constant to safeguard against division by zero.

Overall, the loss function for optimizing the Gaussian scene can be expressed as:
\begin{equation}
    \mathcal{L}_{\text{total}} = \lambda_{\text{rgb}} \mathcal{L}_{\text{RGB}} + \lambda_{\text{pgc}} \mathcal{L}_{\text{PGC}} + \lambda_{\text{dgc}} \mathcal{L}_{\text{DGC}},
\end{equation}
where \(\lambda_{\text{rgb}}\), \(\lambda_{\text{pgc}}\), and \(\lambda_{\text{dgc}}\) are weighting coefficients balancing the contributions of the respective loss terms above.

\section{EXPERIMENT}

\subsection{Experiment Setup}

\subsubsection{Dataset}

This research employs the SCARED dataset \cite{scared2021}, encompassing high-quality videos of fresh porcine cadaver abdominal anatomy captured using a da Vinci Xi endoscope, along with camera trajectories for each frame. The dataset was specifically created to tackle the distinctive challenges of surgical imaging, such as strong directional lighting, non-planar surfaces, and minimal texture, all of which make it more complex than typical natural scenes commonly used in computer vision. For demonstration, 40 consecutive frames were randomly selected from each scene, with 10 frames used for testing and the remaining 30 frames for training. The resolution of each frame was set to $1024\times 1280$. While this setup involves short sequences, the approach is scalable to longer videos for reconstructing more extensive surgical scenes.

\begin{table*}[!ht]
\vspace{3pt} 
\caption{Quantitative results of novel view synthesis and pose estimation for four representative endoscopic scenes. The best outcomes are highlighted in \textcolor{red}{RED}. Pose estimation based on COLMAP is excluded from discussion, as its results are typically considered ground truth.}
\label{Tab:results}
\begin{center}
\renewcommand{\arraystretch}{1.2}
\setlength{\tabcolsep}{4.5mm}{
\begin{tabular}{|c|c|ccc|ccc|}
\hline
                        &                                  & \multicolumn{3}{c|}{Novel View Synthesis}                                                                                             & \multicolumn{3}{c|}{Pose Estimation}                                                                                                 \\ \cline{3-8} 
\multirow{-2}{*}{Scene} & \multirow{-2}{*}{Method}         & \multicolumn{1}{c|}{$\mathbf{PSNR}\uparrow $}                          & \multicolumn{1}{c|}{$\mathbf{SSIM}\uparrow$}                         & $\mathbf{LPIPS}\downarrow$                        & \multicolumn{1}{c|}{$\mathbf{RPE}_{trans}\downarrow$}                   & \multicolumn{1}{c|}{$\mathbf{RPE}_{rot}\downarrow$}                     & $\mathbf{ATE}\downarrow$                          \\ \hline
                        & COLMAP+3DGS                      & \multicolumn{1}{c|}{22.720}                        & \multicolumn{1}{c|}{0.843}                        & 0.327                        & \multicolumn{1}{c|}{-}                            & \multicolumn{1}{c|}{-}                            & -                            \\ \cline{2-8} 
                        & NoPe-NeRF                        & \multicolumn{1}{c|}{34.610}                        & \multicolumn{1}{c|}{0.910}                        & 0.322                        & \multicolumn{1}{c|}{2.377}                        & \multicolumn{1}{c|}{0.435}                        & 0.011                        \\ \cline{2-8} 
                        & CF-3DGS                          & \multicolumn{1}{c|}{33.616}                        & \multicolumn{1}{c|}{0.862}                        & 0.170                        & \multicolumn{1}{c|}{0.478} & \multicolumn{1}{c|}{{\color[HTML]{FE0000} \textbf{0.188}}} & {\color[HTML]{FE0000} \textbf{0.008}} \\ \cline{2-8} 
\multirow{-4}{*}{1}     & \textbf{SurGSplat}               & \multicolumn{1}{c|}{{\color[HTML]{FE0000} \textbf{35.964}}} & \multicolumn{1}{c|}{{\color[HTML]{FE0000} \textbf{0.913}}} & {\color[HTML]{FE0000} \textbf{0.125}} & \multicolumn{1}{c|}{{\color[HTML]{FE0000} \textbf{0.476}}} & \multicolumn{1}{c|}{{\color[HTML]{FE0000} \textbf{0.188}}} & {\color[HTML]{FE0000} \textbf{0.008}} \\ \hline
                        & COLMAP+3DGS                      & \multicolumn{1}{c|}{-}                             & \multicolumn{1}{c|}{-}                            & -                            & \multicolumn{1}{c|}{-}                            & \multicolumn{1}{c|}{-}                            & -                            \\ \cline{2-8} 
                        & NoPe-NeRF                        & \multicolumn{1}{c|}{30.820}                        & \multicolumn{1}{c|}{0.840}                        & 0.316                        & \multicolumn{1}{c|}{2.545}                        & \multicolumn{1}{c|}{0.558}                        & 0.033                        \\ \cline{2-8} 
                        & CF-3DGS                          & \multicolumn{1}{c|}{35.353}                        & \multicolumn{1}{c|}{0.946}                        & 0.079                        & \multicolumn{1}{c|}{{\color[HTML]{FE0000} \textbf{1.012}}} & \multicolumn{1}{c|}{{\color[HTML]{FE0000} \textbf{0.131}}} & {\color[HTML]{FE0000} \textbf{0.026}} \\ \cline{2-8} 
\multirow{-4}{*}{2}     & \textbf{SurGSplat}               & \multicolumn{1}{c|}{{\color[HTML]{FE0000} \textbf{35.422}}} & \multicolumn{1}{c|}{{\color[HTML]{FE0000} \textbf{0.947}}} & {\color[HTML]{FE0000} \textbf{0.075}} & \multicolumn{1}{c|}{1.013}                        & \multicolumn{1}{c|}{0.132}                        & {\color[HTML]{FE0000} \textbf{0.026}} \\ \hline
                        & COLMAP+3DGS                      & \multicolumn{1}{c|}{-}                             & \multicolumn{1}{c|}{-}                            & -                            & \multicolumn{1}{c|}{-}                            & \multicolumn{1}{c|}{-}                            & -                            \\ \cline{2-8} 
                        & NoPe-NeRF                        & \multicolumn{1}{c|}{28.251}                        & \multicolumn{1}{c|}{0.758}                        & 0.401                        & \multicolumn{1}{c|}{4.059}                        & \multicolumn{1}{c|}{0.531}                        & 0.056                        \\ \cline{2-8} 
                        & CF-3DGS                          & \multicolumn{1}{c|}{31.218}                        & \multicolumn{1}{c|}{0.901}                        & 0.128                        & \multicolumn{1}{c|}{1.029}                        & \multicolumn{1}{c|}{{\color[HTML]{FE0000} \textbf{0.262}}} & {\color[HTML]{FE0000} \textbf{0.033}} \\ \cline{2-8} 
\multirow{-4}{*}{3}     & \textbf{SurGSplat}               & \multicolumn{1}{c|}{{\color[HTML]{FE0000} \textbf{31.299}}} & \multicolumn{1}{c|}{{\color[HTML]{FE0000} \textbf{0.902}}} & {\color[HTML]{FE0000} \textbf{0.125}} & \multicolumn{1}{c|}{{\color[HTML]{FE0000} \textbf{1.028}}} & \multicolumn{1}{c|}{{\color[HTML]{FE0000} \textbf{0.262}}} & {\color[HTML]{FE0000} \textbf{0.033}} \\ \hline
                        & \multicolumn{1}{l|}{COLMAP+3DGS} & \multicolumn{1}{c|}{-}                             & \multicolumn{1}{c|}{-}                            & -                            & \multicolumn{1}{c|}{-}                            & \multicolumn{1}{c|}{-}                            & -                            \\ \cline{2-8} 
                        & NoPe-NeRF                        & \multicolumn{1}{c|}{32.195}                        & \multicolumn{1}{c|}{0.872}                        & 0.301                        & \multicolumn{1}{c|}{4.124}                        & \multicolumn{1}{c|}{0.596}                        & 0.041                        \\ \cline{2-8} 
                        & CF-3DGS                          & \multicolumn{1}{c|}{36.611}                        & \multicolumn{1}{c|}{0.957}                        & 0.080                        & \multicolumn{1}{c|}{\color[HTML]{FE0000} \textbf{1.525}}                        & \multicolumn{1}{c|}{0.083}                        & {\color[HTML]{FE0000} \textbf{0.039}} \\ \cline{2-8} 
\multirow{-4}{*}{4}     & \textbf{SurGSplat}               & \multicolumn{1}{c|}{{\color[HTML]{FE0000} \textbf{36.882}}} & \multicolumn{1}{c|}{{\color[HTML]{FE0000} \textbf{0.958}}} & {\color[HTML]{FE0000} \textbf{0.079}} & \multicolumn{1}{c|}{1.531} & \multicolumn{1}{c|}{{\color[HTML]{FE0000} \textbf{0.082}}} & {\color[HTML]{FE0000} \textbf{0.039}} \\ \hline
\end{tabular}}
\end{center}
\end{table*}

\subsubsection{Implementation Details}

Experimental evaluations were carried out within the PyTorch framework on a single NVIDIA RTX 4090 GPU with 24GB of memory. For monocular depth estimation, we utilized the Depth Anything \cite{depthanything2024} model to compute depth maps $\{D_t\}, t \in (0, N-1)$, while matching points between consecutive frames $I_{t-1}$ and $I_t$ were identified using the pre-trained RoMa \cite{roma2023} model. During optimization, the interval for adding new frames was synchronized with the interval for point densification, promoting incremental growth of the reconstructed scene. Furthermore, opacity values were reset consistently throughout the training process to ensure the effective incorporation of new frames into the Gaussian model derived from previously observed frames. Unless explicitly stated, all other parameter settings remained consistent with the original 3DGS configuration.

\subsection{Comparative Experiment Settings}
To validate the superiority of our proposed algorithm, we performed comparative experiments against several baseline methods, including the traditional 3DGS model initialized with COLMAP, as well as two SfM-free approaches, NoPe-NeRF and CF-3DGS.

For novel view synthesis (NVS) evaluation, we adopted standard image quality metrics: $\mathbf{PSNR}$, $\mathbf{SSIM}$ \cite{ssim2004}, and $\mathbf{LPIPS}$ \cite{lpips2018}, providing a comprehensive assessment of the visual fidelity and perceptual quality of the generated images. Additionally, camera pose estimation was evaluated using the Absolute Trajectory Error ($\mathbf{ATE}$) \cite{nopenerf2023}, along with the Relative Pose Error for translation ($\mathbf{RPE}_{trans}$) and rotation ($\mathbf{RPE}_{rot}$), which measure frame-to-frame pose accuracy to capture finer trajectory deviations from ground truth poses.

\section{RESULTS AND DISCUSSION}

We carried out rigorous evaluations of our proposed SurGSplat method on the SCARED dataset, highlighting its outstanding performance in NVS and pose estimation. Fig.\ref{Fig:3} presents novel synthesized images from four representative scenes, highlighting the effectiveness of our approach. Notably, NeRF-based methods produced blurry renderings with significant detail loss, and their inaccurate camera poses posed potential risks for downstream tasks such as intraoperative navigation and lesion segmentation. The inherent challenges of endoscopic imagery, including simple textures and complex structures, led to reconstruction failures in multiple scenes for COLMAP. Even when using basic Gaussian Splatting, the novel-rendered view of Scene 1 exhibited incomplete reconstruction, with a portion of the structures missing. Additionally, CF-3DGS encountered noticeable artifacts during the reconstruction of Scene 1 and Scene 3, likely due to inaccuracies in feature point matching. These misalignments, particularly in scenes with sparse textures or repetitive patterns, hindered the reconstruction process. In contrast, our method demonstrated significantly improved rendering quality and robust scene reconstruction.

Quantitative results, shown in Table \ref{Tab:results}, further support these observations. Subtle differences, imperceptible to the naked eye, were reflected in the metrics, where the relative pose estimation method achieved the lowest average trajectory error, indicating superior accuracy. Moreover, SurGSplat surpassed all other methods in NVS metrics and showed outstanding performance in trajectory accuracy, underscoring its capability to generate high-fidelity images and accurate pose estimations. Importantly, our algorithm enables the reconstruction of fine anatomical details, such as blood vessels, which are critical for clinical decision-making during surgery. By providing enhanced visual clarity and precision, SurGSplat supports surgeons in navigating complex anatomical structures, improving the safety and efficacy of intraoperative procedures. These results highlight not only the clinical applicability of our method but also its necessity in addressing the unique challenges of endoscopic navigation, paving the way for more reliable and precise surgical interventions.

Despite the promising results achieved by our SurGSplat method, there remain areas for future improvement. Currently, our algorithm is specifically designed for dense endoscopic video data to reconstruct surgical scenes with high fidelity. However, in clinical practice, there are scenarios where only sparse viewpoints are available, such as during preoperative planning or when navigating constrained anatomical regions. Future extensions of our method will focus on adapting it for sparse-view reconstructions, which would broaden its applicability to a wider range of clinical workflows.

\section{CONCLUSION}

This paper proposes \textbf{SurGSplat}, an advanced framework developed to tackle the challenges of 3D reconstruction in endoscopic environments. By the local refinement and integrating geometric consistency constraints into the progressive optimization of 3D Gaussian Splatting, the method enables precise pose estimation and high-resolution scene reconstruction, eliminating the dependence on conventional SfM techniques. Comprehensive evaluations demonstrate the robustness and accuracy of SurGSplat, highlighting its significant potential to improve the efficiency and precision of surgical navigation systems.

\addtolength{\textheight}{-12cm}   








\bibliographystyle{IEEEtran}
\bibliography{ref}
\end{sloppypar}
\end{document}